# Sum frequency generation spectroscopy of the attachment disc of a spider

Yue Zhao [1], Lin Liang [1], Yanrong Li [1], Khuat Thi Thu Hien [1], Goro Mizutani [1] *, and Harvey N. Rutt [2]

[1] School of Materials Science, Japan Advanced Institute of Science and Technology, 1-1 Asahidai, Nomi 923-1292, Japan.
[2] School of Electronic and Computer Science, University of Southampton, Southampton, SO17 1BJ, UK.

**Abstract**
The pyriform silk of the attachment disc of a spider was studied using infrared-visible vibrational sum frequency generation (SFG) spectroscopy. The spider can attach dragline and radial lines to many kinds of substrates in nature (concrete, alloy, metal, glass, plant branches, leaves, etc.) with the attachment disc. The adhesion can bear the spider's own weight, and resist the wind on its orb web. From our SFG spectroscopy study, the NH group of arginine side chain and/or $NH_2$ group of arginine and glutamine side chain in the amino acid sequence of the attachment silk proteins are suggested to be oriented in the disc. It was inferred from the observed doublet SFG peaks at around 3300 cm$^{-1}$ that the oriented peptide contains two kinds of structures.

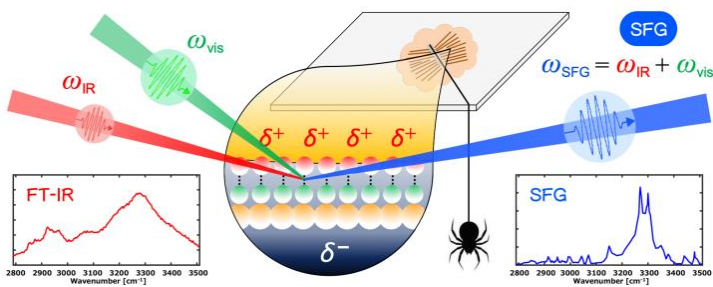

* Correspondence to Goro Mizutani
   Address: 1-1 Asahidai, Nomi 923-1292, Japan.
   Phone number: +81-761-51-1521
   Fax number: +81-761-51-1149
   E-mail: mizutani@jaist.ac.jp

## 1. Introduction

The orb-weaving spider can secrete seven types of silk for various uses by its seven types of secretory glands. The spider's dragline for vertical movement and the radial lines for orb-web construction are secreted from the large ampullate gland. On the other hand, the pyriform gland secretion is used to glue a dragline or a radial line onto surfaces [1–4]. Pyriform silk and glue are extruded from pyriform gland [5]. The pyriform gland is composed of two kinds of secretory cells [5]. One is located in the distal half of the glands, and produces pyriform spidroins with finely fibrillar proteinic granules through an extensive rough endoplasmic reticulum; another one is located in the proximal half of the gland, and secretes a glue-like component containing protein with a highly organized structure and unidentified carbohydrate [5]. In the proximal half of the gland, the carbohydrate component may be added to the protein [5].

The glue from the pyriform gland is an analogous liquid [6]. A stable cement-like attachment disc is formed in less than a second in a natural environment [7], and sticks the dragline and radial lines to the surface of materials [8]. The structure made of pyriform silk and cement-like glue for fixing the dragline and the radial line to the materials is called an attachment disc. Hundreds of spinnerets in the abdomen of the spider are brought into direct contact with the material's surface, and a large amount of pyriform silk is discharged from the spinnerets while being rubbed against the surface to form an attachment disc [9–12]. When a spider secretes attachment disc silk, its spinnerets are said to rub back-and-forth on parallel straight lines [12]. As shown in Figs. 1(b) and (c), the attachment disc of pyriform silk looks like a set of staple-pins and it fixes the dragline to the substrate.

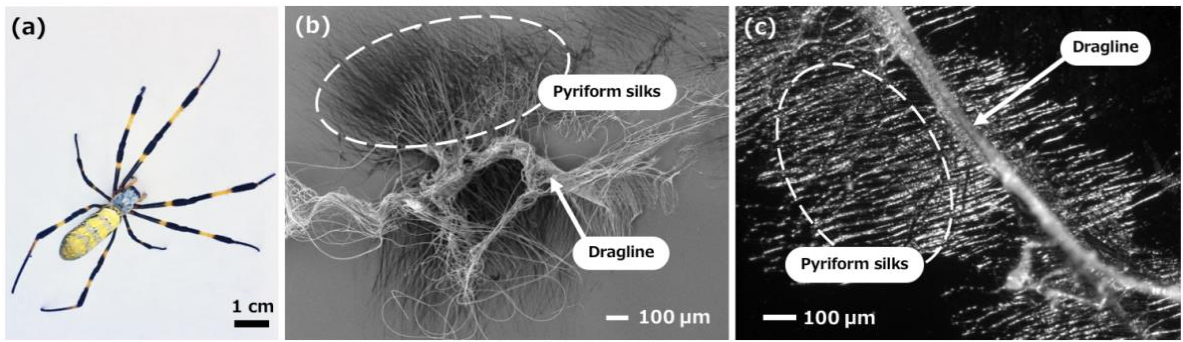

**Figure 1.** (a) Photograph of the spider (*N. clavata*) featured in this study. (b) Scanning electron microscopy (SEM, HITACHI TM3030Plus, acceleration voltage: 15 kV) image of an attachment disc of the spider. (c) Optical microscopy image of an attachment disc of the spider.

The attachment disc can apply a strong adhesive force to the surface of almost any material in nature. On the other hand, it has been reported that its adhesion strength or weakness depends greatly on the material [13,14]. However, the attachment disc adhesion strength to polytetrafluoroethylene (PTFE) is high in spite of its very low surface free energy [13]. On the other hand, the attachment disc cannot adhere to highly hydrophobic substrate surfaces, such as a glass treated with dichlorodimethylsilane [7]. Besides this example, the mechanism of its adhesion remains unclear in many aspects.

Pyriform silk is mainly composed of proteins called pyriform spidroin 1 (PySp1) and pyriform spidroin 2 (PySp2) [6,8,15]. PySp1 contains block modules with a high degree of polarity and charge [6]. On the molecular scale, the amino acid sequences of pyriform silks may interact strongly with the cement-like glue due to a high content of polar side chains. Pyriform silks are not directly involved in the interaction with the substrate [7], but must be firmly attached to the cement-like glue. We have found in the past that an anisotropic secondary structure in the highly oriented spidroin (e.g., dragline or radial line) exhibits a second-order nonlinear optical response [16,17]. In this study, in order to get a further hint of the structure of the pyriform silk, vibrational sum frequency generation (SFG) spectroscopy was used to investigate the physically anisotropic structure and the chemical species of the attachment disc. Since SFG occurs only in non-centrosymmetric structures, it can selectively observe oriented molecular components. Hence orientation of asymmetric molecular species can be analyzed by SFG. Since the SFG response is very weak, one must have as much sample material as possible in the probe laser spot, in order to get good S/N data. From this point of view also, attachment discs are easier to study than the other types of spider silk, since they are secreted in disc shapes and not in thin line shapes.

With the development of optical engineering, SFG vibrational spectroscopy has been widely used for the study of protein structure and molecular dynamics at various *in situ* surfaces and interfaces for the last 20 years [18–42]. The secondary structure of proteins can be analyzed by examining chiral NH stretching vibration and chiral amide I using chiral SFG [23,41]. On the other hand, our study uses the 'achiral' SFG method and it cannot analyze the secondary structures. While it is important to investigate the secondary structure of proteins in spider attachments, there is no SFG study on attachment in the past. Hence this study is focused just on the second-order nonlinear response in molecular species. Here, the 'achiral' SFG method not only lets us assign the molecular species but also selectively detects non-centrosymmetric bulk structures including the secondary structure of proteins.

The cement-like glue has been reported to be protein rich in acidic groups and associated with a carbohydrate component [5]. A beads-on-a-string (BOAS) structure similar to the sticky aggregate glue droplets of orb web for catching preys is formed occasionally when the pyriform gland secretions are not in contact with the substrate [7]. Singla *et al.* [43] reported the SFG spectrum of the aggregate glue droplets of a spider orb web. Aggregate glue droplet is composed of glycoproteins (ASG-1 and ASG-2) [44–49], low-molecular-mass compounds (LMMCs) [50–55], inorganic salts [51,53,54], and bound water [43,56]. The attachment disc and the aggregate glue droplet have two similarities: the BOAS structure and the protein associated with a carbohydrate component. The attachment discs observed in our study should contain components in common with the aggregate glue droplet sample of Singla *et al.* [43]. On the other hand, our study also revealed different components from that in aggregate glue droplets.

## 2. Experimental Section

The samples consisted of attachment discs of *N. clavata,* or a spider shown in Fig. 1(a). The samples were collected 3-5 days before observation. The attachment discs were adhered to silicon wafer substrates by the spider. When we collected the samples, several silicon wafers (2 cm x 2 cm) were stuck on a sufficiently large sheet of paper. Spiders were captured on the JAIST campus (coordinates (WGS 84): E 136.5955, N 36.4451) in Ishikawa Prefecture. When a spider walked on the paper, it sometimes secreted an attachment disc under its belly. While secreting the pyriform disc, the spider stopped and shook its belly part to left and right. We gently encouraged the spider to walk on the Si wafers and to secrete attachment discs on to the wafers. The detail of the sample collection is shown in the *Supporting data* (see Video S1).

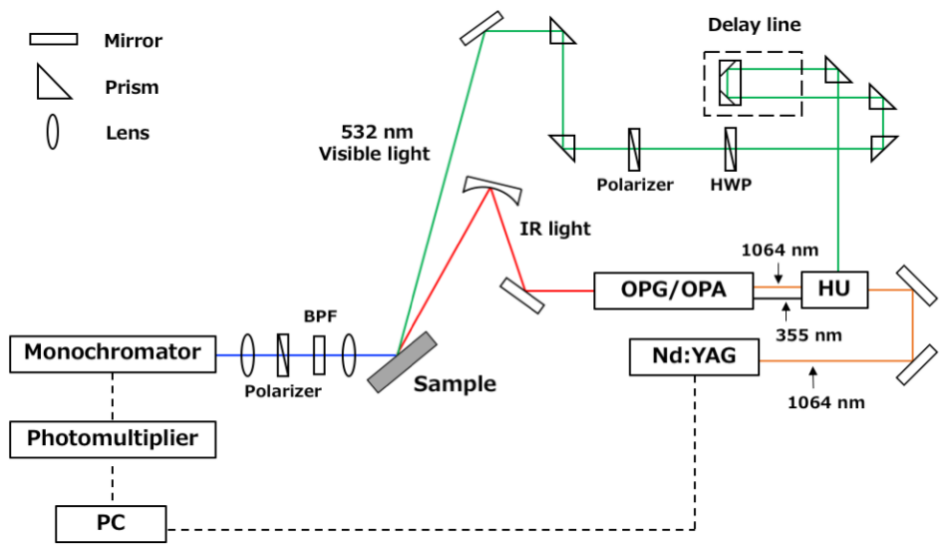

**Figure 2.** The optical setup of our SFG spectroscopy system. BPF: Band pass filter. HU: Harmonic unit. OPG/OPA: Optical parametric generation/amplifier. HWP: Half wave plate.

Figure 2 shows the optical setup of our SFG measurement system. In this study, we used doubled-frequency light pulses at a visible wavelength of 532 nm with the pulse time width set to 30 ps and the repetition rate of 10 Hz generated by a harmonic unit (EKSPLA HMPL/SH/TH/FH 183) excited by a mode-locked Nd:YAG laser (EKSPLA PL2143) system. Output from an optical parametric generator and amplifier system (OPG/OPA) (EKSPLA P6401DFG2-18) driven by the same Nd:YAG laser was used as the wavelength-tunable infrared (IR) light. The SFG signal was generated in the reflection direction from the sample by the IR and visible light pulses. It was then introduced into a monochromator (Nihon Koken Kogyo Co., Ltd., SG-100) and finally detected and recorded by a photomultiplier (HAMAMATSU R585) and a gated integrator system. The resolution of the monochromator was set as 13.6 nm in the SFG measurement. The wavenumber resolution was determined by the spectral width of the IR output from our OPG/OPA and it was 3 $cm^{-1}$. Both visible and IR light beam spots were focused on the same area of 2 $mm^2$ on the sample surface. The energies of the visible and IR light were 35 μJ/pulse and 25-115 μJ/pulse, respectively. The corresponding flux of the visible and IR pulses were 1.75 $mJ/cm^2$·pulse and 1.25-5.75 $mJ/cm^2$·pulse, and the peak power densities on the samples were 58.3 $TW/cm^2$ and 41.7-192 $TW/cm^2$, respectively. The polarization configurations PPP and SSP were used. Here the three capital letters denote the polarizations of the sum frequency output, visible input, and IR input pulses in that order. The maximum output energy of the IR light depended on the output wavelength of our OPG/OPA. The measured SFG signal was normalized by that from a GaAs(001) wafer under the same conditions. After the SFG measurement no damage of either nanometer or micrometer scale was detected on the sample using an optical microscope or an SEM. Additionally, since there is no time dependency of SFG, the damage was considered to be of negligible impact.

The measurement step of the Fourier-transform Infrared Spectrometer (FT-IR) (Perkin Elmer, Spectrum 100) was 1 $cm^{-1}$ and the spectral resolution was 4 $cm^{-1}$. The attachment disc samples were separated from the Si substrates and measured on their own by attenuated total reflection (ATR).

## 3. Results

Figure 3 shows an FT-IR absorption spectrum of an attachment disc in the wavenumber region of 900-1800 cm$^{-1}$. Three absorption bands can be seen at 1200-1260 cm$^{-1}$, 1510-1580 cm$^{-1}$, and 1600-1700 cm$^{-1}$. These are the characteristic bands of the peptide group named amide III, amide II, and amide I, as indicated in the figure. Two absorption peaks are seen at 1167 cm$^{-1}$ and 1454 cm$^{-1}$ and are supposed to be the bands of β-poly-*L*-alanine (β-PAla) [57]. These results are consistent with the IR absorption spectrum of the dragline of the spider already reported [57]. Two more absorption peaks are seen at 1050 cm$^{-1}$ and 1400 cm$^{-1}$. They are ascribed to serine [58] and/or lipids [59,60].

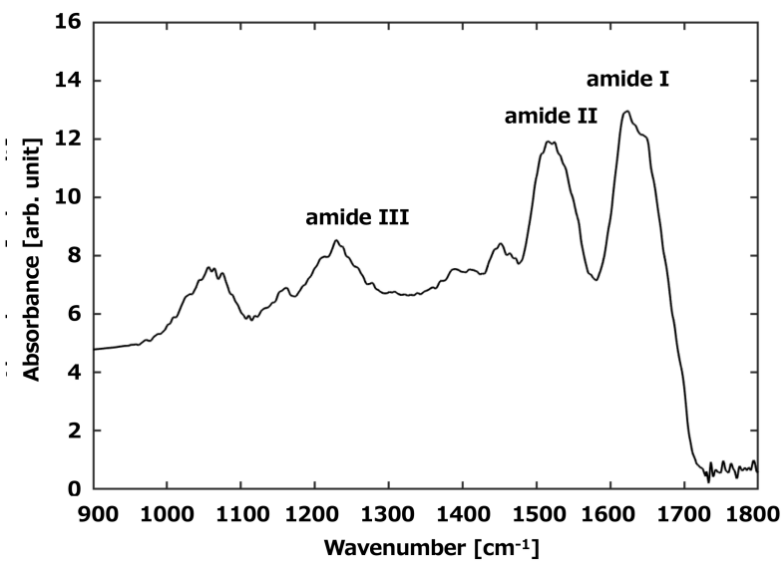

**Figure 3.** FT-IR spectrum of the 900-1800 cm$^{-1}$ region of an attachment disc of a spider (*N. clavata*). Bands attributed to amide I, II and III are marked.

Figure 4 shows an FT-IR absorption spectrum and SFG spectra of an attachment disc in the wavenumber region of 2800-3500 cm$^{-1}$. The wavenumber resolution of the SFG spectrum is about 10 cm$^{-1}$ after a Savitzky-Golay smoothing treatment [61] with three smoothing points. The incident plane of the two excitation light fields is parallel to the pyriform silk fiber axis of the attachment disc. The polarization configuration are PPP and SSP for the middle and bottom spectra, respectively. We obtained SFG spectra at seven different positions on four different attachment disc samples with good reproducibility. The IR absorption spectrum shows two wide vibrational bands at 2850-3000 cm$^{-1}$ and 3100-3500 cm$^{-1}$. The IR spectrum at 2800-3500 cm$^{-1}$ in Fig. 4(a) is consistent with the IR spectrum of the spider dragline reported by Papadopoulos *et al.* [57].

The wavenumbers of the 13 detected peaks in the SFG spectra are shown in Fig. 4(b). The bands below 3000 cm$^{-1}$ are assigned to CH$_n$ vibrations. 3044 cm$^{-1}$ and 3070 cm$^{-1}$ peaks are assigned to the CH stretching of the aromatic ring. 3270 cm$^{-1}$ and 3300 cm$^{-1}$ are assigned to NH of β-sheet [62–64] and α-helix [65–69], respectively. The assignment of the SFG bands in Fig. 4(b) based on the literature is shown in Table 1. In the polarization configuration of SSP, the relative intensity of the peak at 3150 cm$^{-1}$ is slightly stronger than that in the PPP configuration.

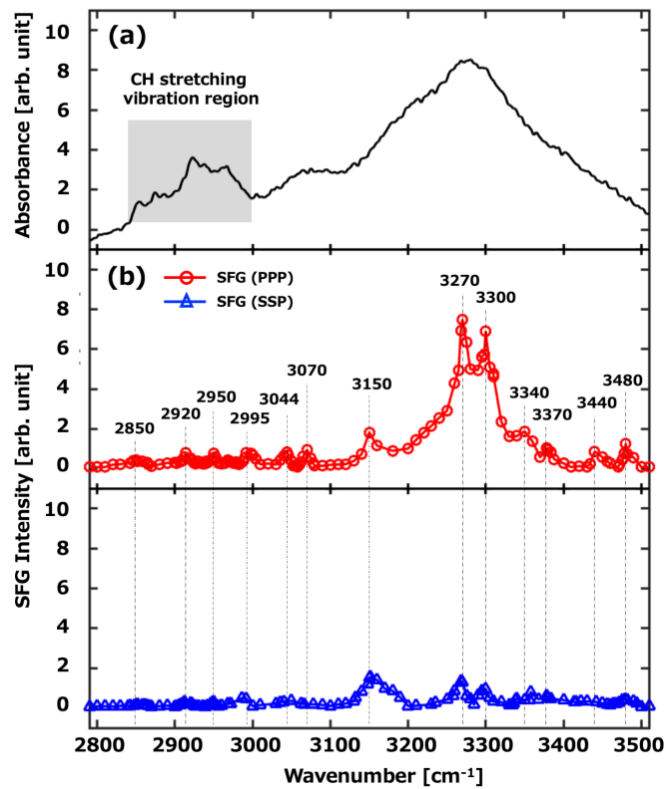

**Figure 4.** (a) FT-IR and (b) SFG spectra of the 2800-3500 cm$^{-1}$ region of an attachment disc of a spider (*N. clavata*). The solid lines of the SFG spectra are the result of Savitzky-Golay smoothing. The wavenumber resolution is ~10 cm$^{-1}$ after the smoothing.

The SFG peaks were detected only when the incident plane and the fiber axes of the attachment disc silk were parallel to one other. When the fiber axes of the attachment disc silk were perpendicular to the incident plane, the incident light was scattered strongly by the attachment disc and did not enter the detector in the regular reflection direction. The linear optical reflectance of the attachment disc on the silicon substrate increased linearly as a function of the angle between the silk fiber axis and the incident plane. The reflectance is at the maximum when the fiber axis of the silk is parallel to the incident surface (see Figure S1). We tried to measure the chiral SFG component by using a combination of SPP and PSP polarization [70], but the signal was below the noise level.

## 4. Discussion
### 4.1. Assignment of the observed peaks

In this study, our main concern is the peaks at 3150 cm$^{-1}$, 3270 cm$^{-1}$ and 3300 cm$^{-1}$. Here, before going to these peaks, we check the observed peaks in the FT-IR and SFG spectra. Figures 3 and 4(a) show IR vibration bands in the 900-1800 cm$^{-1}$ and 2800-3600 cm$^{-1}$ regions. FT-IR data of 1800-2800 cm$^{-1}$ is shown in Figure S2 in the *Supporting data* file. The infrared absorption spectrum of the attachment disc silk in Fig. 3 including the IR vibration bands of amides I, II and III, is quite similar to that of the spider's dragline already reported [57]. This is because both of them consist of proteins.

Figure 4(b) shows SFG bands in the 2800-3600 cm$^{-1}$ region. The non-zero second-order nonlinear optical response of the attachment disc silk was confirmed by a second harmonic generation (SHG) microscopy (see Figure S3 in the *Supporting data*). Figure 4 shows several peaks in the SFG and IR spectra. In our experimental process, bare silicon substrates give SFG signals at the noise level in the CH, NH and OH stretching region at 2800-3600 cm$^{-1}$. Si is famous for having a strong SHG response from its surface dangling bonds, but SHG and SFG become very weak if there is a native oxide layer on the surface [71]. The fact that the SFG from bare Si is negligible in our experimental condition further means that SFG contribution from environmental contamination is also at our noise level. Additionally, our previous paper [16] reported the existence of "SHG" signals from the radial lines and draglines of spider silk but not from the spiral lines. If the contamination with non-centrosymmetric structures in our sample preparation process contributed to the second order optical nonlinearity, there should be SHG signals from all the types of spider silk. Hence, we can say that the effects of contamination with non-centrosymmetric structures can be excluded from our sample preparation conditions in our later discussion.

**Table 1.** Tentative assignment of the 2800-3500 cm$^{-1}$ region of the vibrational SFG peaks of the attachment disc of the spider (*N. clavata*).

| attachment disc SFG (cm$^{-1}$) | Possible assignment |
|---|---|
| 2850 | $\nu_{ss}(CH_2)$ [72–75] |
| 2920 | $\nu_{as}(CH_2)$ [75] |
| 2950 | Fermi-resonance of $\nu_{as}(CH_3)$ [76] or $OCH_3$ [77–79] |
| 2995 | $\nu_{as}(CH)$ [80] |
| 3044 | $\nu(CH)$ aromatic ring [81–84] |
| 3070 | $\nu(CH)$ aromatic ring [81,84–86] |
| 3150 | NH of Arg side chain [87], Fermi-resonance of $\nu_{ss}(NH_2)$ [88] |
| 3270 | β-sheet [62–64] |
| 3300 | α-helix [65–69] |
| 3340 | NH [67] |
| 3370 | OH [89,90], $\nu_{ss}(NH)$ [91] |
| 3440 | OH of hydroxyl group [92–94], NH [66,67,95,96] |
| 3480 | OH [97], NH [66] |

$\nu$: stretching vibration, $\nu_{ss}$: symmetric stretching, $\nu_{as}$: asymmetric stretching

As shown in Table 1, four peaks of the CH stretching vibration region (2850-3000 cm$^{-1}$) in the FT-IR spectrum in Fig. 4(a) are assigned to: CH$_2$ symmetric stretching [72–75] at 2850 cm$^{-1}$, CH$_2$ asymmetric stretching [75] at 2916-2924 cm$^{-1}$, Fermi-resonance of $\nu_{as}(CH_3)$ [76] or OCH$_3$ [77–79] at 2950 cm$^{-1}$, and CH asymmetric stretching [80] at 2995 cm$^{-1}$. As the origin of the aromatic ring signal, phenylalanine and tyrosine are considered among the amino acids constituting pyriform spidroin [8,98].

The NH stretching vibration usually serves as a probe to investigate the structural and functional properties of proteins [66,68]. Fermi resonance by the NH stretching vibration and the amide II harmonic leads to bands [66] at 3100-3500 cm$^{-1}$. The peak of 3150 cm$^{-1}$ is discussed in detail in section *4.4*. The SFG peaks at 3270 cm$^{-1}$ and in the 3300-3360 cm$^{-1}$ region in Fig. 4(b) are typical modes in NH stretching vibration [62–64,66–69]. At circa 1630 cm$^{-1}$ in the FT-IR spectrum in Fig. 3, we see a typical amide I peak [99]. For the SFG peaks between 3270 and 3300 cm$^{-1}$, the intensity for polarization combination of PPP is significantly stronger than that of SSP. Since there is non-zero second order nonlinearity arising from NH bonds, it is suggested that the second order nonlinearity was caused by oriented amino acid residues. The doublet peaks of 3270 and 3300 cm$^{-1}$ are discussed in detail in section *4.3*. The SFG peak at 3440 cm$^{-1}$ is near the frequency 3480 cm$^{-1}$ in the hydroxyl vibration region of protein [92–94,100,101]. However, the assignment of the peaks between 3370 cm$^{-1}$ and 3480 cm$^{-1}$ is difficult.

### *4.2. Comparison of the SFG spectral data in this study with those of the aggregate glue droplet on an orb web reported by Singla et al.* [43]

Here we compare the SFG spectra of the attachment disc in this study with those of the aggregate glue droplet of the orb web and its LMMCs extract by Singla *et al.* [43]. The common features of the spectra shared by the attachment disc, the aggregate glue droplet and the LMMCs are as below.
1) The peaks of CH$_n$ at 2800-3000 cm$^{-1}$ are seen.
2) The peak centered at 3270 cm$^{-1}$ are seen for the attachment disc and the glycoproteins in aggregate glue droplet.
3) The SFG spectra of LMMCs [43] show a peak near 3150 cm$^{-1}$ for the SSP polarization combination. The enhancement of the SFG peak intensity at 3150 cm$^{-1}$ for SSP polarization combination is also seen for the attachment disc in Fig. 4(b).

Differences are in the following.
 i) No peak was observed around 3270 cm$^{-1}$ for LMMCs by Singla *et al.* [29].
 ii) At 3270 cm$^{-1}$ the SFG spectra of aggregate glue droplet by Singla *et al.* shows one broad peak [29], while the attachment disc in this study shows doublet peaks at 3270 cm$^{-1}$ and 3300 cm$^{-1}$.
 iii) The peaks at 3150 cm$^{-1}$ in the SFG spectra of the attachment disc in this study are prominent, but not in that of aggregate glue droplet.

From the common features of 1) and 2), we guess that both the attachment disc and the aggregate glue droplet contain hydrocarbons and peptide chains. As it is well known, the early secretions of the component associated with the carbohydrate of the cement-like glue of the attachment disc appear in cells with the Golgi apparatus in the proximal part of the pyriform gland [5]. Since most glycosylation reactions occur in the Golgi apparatus [102], the component associated with the carbohydrate in the cement-like glue of the attachment disc may be a glycoprotein.

From the difference point i), the SFG spectrum of the LMMCs extract by Singla *et al.* shows no NH peak [43] and so there is no oriented peptide bond in LMMCs. In fact, the LMMCs in aggregate glue droplet consist of *γ*-aminobutyramide, *N*-acetyltaurine, choline, betaine, isethionic acid and pyrrolidone [50–55], and so there can be no peptide bonds in LMMCs. On the other hand, the common feature 3) suggests the existence of LMMCs in the cement-like glue of the attachment disk like in aggregate glue droplet. Concerning the difference point ii), the splitting of the peak is discussed in detail in section *4.3*. The peak at 3150 cm$^{-1}$ mentioned in the difference point iii) can be assigned either as LMMCs, glutamine, or arginine in spidroin, as it will be discussed in section *4.4*.

Summarizing this section, the SFG spectra show that glycoproteins may exist both in the attachment disc and aggregate glue droplet. On the other hand, since there are differences in the SFG spectra, the attachment disc contains different peptides aggregation from that in the aggregate glue droplet.

### *4.3. Discussion on the origin of the splitting of 3270 and 3300 cm$^{-1}$ peaks*

The infrared absorption frequency of secondary amide in the solid state depends on the type of hydrogen bonding. In some cases of dipeptides, polypeptides, and proteins, two or multiple absorption peaks are found [66]. Since the main NH absorption of secondary amides occurs near 3270 cm$^{-1}$ in the solid state [66,103–105], the 3270 cm$^{-1}$ peak in Fig. 4(b) can be assigned to the secondary amides. The splitting of the 3270 and 3300 cm$^{-1}$ peaks is considered to be due to two kinds of structures in (1) α-helical and/or (2) β-sheet structures. For discussing candidate (1), we refer to the results of Lee *et al.* [65]. Experimentally, oriented and unoriented α-helical poly (*L*-alanine) (β-PLA) show doublet structures at amide A peak in IR [65] and polarized Raman spectra [106]. This amide A doublet was suggested to arise from two different structures, and the higher frequency component was assigned to the standard α-helical structure in the peptide chains [65]. Thus, in our result, the peak at 3300 cm$^{-1}$ in Fig. 4(b) can be assigned to the α-helical structure. In Lee *et al.*'s paper [65], the lower frequency peak was not assigned, but was suggested to originate from another peptide structure. Thus, the peptides detected by SFG in this study may consist of two structures.

Since 3270 cm$^{-1}$ peak in the FT-IR spectrum is frequently assigned to a hydrogen bond in the β-sheet [62–64], the β-sheet can be the candidate origin (2) of 3270 cm$^{-1}$ peak. The peak splitting may originate from two kinds of structures in the peptide chain of β-sheet. The band between 3270 and 3310 cm$^{-1}$ is normally modified by two factors: transition dipole coupling (TDC) and Fermi resonance [107]. The resonance frequency in NH mode depends on the strength of the NH⋯O=C hydrogen bond, and the farther the distance of H⋯O, the higher the resonance frequency [107]. Therefore, the NH⋯O=C hydrogen bond strength related to the SFG peak at 3300 cm$^{-1}$ should be weaker than that at 3270 cm$^{-1}$.

Here, we note that the peak at 3270 cm$^{-1}$ was observed in a broad peak in FT-IR, but was observed in a doublet by SFG. In FT-IR vibrations of all the molecular bonds are observed regardless of the molecular orientation. In SFG non-centrosymmetric structures or oriented asymmetric molecules are selectively observed. Namely, in Fig. 4(b) two types of oriented peptides were selectively observed by SFG from differently oriented peptides with various types of structures. On the other hand, only one peptide structure was observed in aggregate glue droplets by Singla *et al*. [43]

### *4.4. Discussion on the origin of the 3150 cm$^{-1}$ peak and its polarization dependence in the SFG spectra*

Figure 4(b) shows peaks in the SFG spectra at 3150 cm$^{-1}$. There are five candidate origins of this peak: 1) NH group of arginine side chains; 2) NH$_2$ group of glutamine and/or arginine side chains; 3) LMMCs; 4) reorganized molecular bonding between the cement-like glue and pyriform spidroin and the 5) cyclic lactam with ring sizes of seven or less.

First, the 3150 cm$^{-1}$ peak may arise from the NH groups or NH$_2$ groups in the side chains of amino acid. NH group is contained in the side chain of arginine, histidine, and tryptophan among the 20 kinds of amino acids constituting natural protein. Pyriform spidroin forming the spider silk contains ~2.6% arginine (R), ~0.05% histidine (H), and does not contain tryptophan (W) [8,98]. One histidine is present in the amino acid sequences of piriform silk proteins of the *A. trifasciata* spider, but not in *N. clavate* [98] studied in this work. The 3150 cm$^{-1}$ peak is characteristic of NH stretch in the side chain of arginine [87,108]. So here, as candidate 1), the SFG peak of 3150 cm$^{-1}$ can be assigned as NH group of arginine side chains.

On the other hand, the 3150 cm$^{-1}$ peak is frequently assigned as the Fermi-resonance splitting of $\nu_{ss}(NH_2)$ [88]. $NH_2$ group is contained in the side chain of aspartic acid, glutamine, arginine, asparagine, and lysine among the 20 kinds of amino acids constituting natural protein. Pyriform spidroin forming the spider silk contains ~15.5% glutamine (Q), ~2.6% arginine (R), ~2.3% asparagine (N), and ~0.5% lysine (K) [8,98]. Furthermore, cDNA analysis has identified repetitive motifs (e.g., QQASVSQS and QQSSLAQS) made of 8 amino acids in the pyriform spidroin, and these motifs contain a large amount of glutamine [98]. Among these amino acid molecules only glutamine and arginine have an vibrational resonance peak near 3150 cm$^{-1}$ [109], while asparagine [110], and lysine [111] does not. So here, as candidate 2), the SFG peak of 3150 cm$^{-1}$ can be assigned as $NH_2$ group of glutamine and arginine side chains.

As explained in the common feature 3) in section *4.2*, the SFG peaks at 3150 cm$^{-1}$ of the attachment disc and LMMCs of aggregate glue droplet shared similar polarization dependence. The relative intensity of the SFG peak at the 3150 cm$^{-1}$ SSP polarization combination of the attachment disc is strong, but that of the LMMCs is weak. The possibility that the SFG peak of 3150 cm$^{-1}$ on the attachment disc is derived from LMMCs cannot be ruled out as candidate 3). However, since 3150 cm$^{-1}$ of LMMCs is very weak, the peak of 3150 cm$^{-1}$ in Fig. 4(b) should be dominated by contributions from other molecules than LMMCs.

In a report of SFG spectra by Dreesen *et al.* [112] on the binding of avidin and biocytin, the interaction between the two substances created a specific order in the molecule. Two new $\nu(CH)$ vibrations centered on 3050 cm$^{-1}$ and $\nu(NH)$ centered on 3150 cm$^{-1}$ were obtained. They were not originally present in avidin or biotin when they were separate [112]. The reorganization of the biocytin conformation after bonding to avidin eliminated the vibrations in the original $CH_2$ chain of biocytin [112]. The NH vibration at 3150 cm$^{-1}$ can be assigned to avidin, biocytin, or both [112]. This result of Dreesen *et al.* [112] can explain the SFG peak 3150 cm$^{-1}$ for the attachment disc in Fig. 4. In the attachment disk, the pyriform spidroin silks are thought to be bound to the cement-like glue [7]. New vibration modes of 3044 cm$^{-1}$ $\nu(CH)$ and 3150 cm$^{-1}$ $\nu(NH)$ may have been created by the interaction between the carbohydrate component of cement-like glue and pyriform spidroin (candidate 4)). These two peaks were not in the SFG spectrum of aggregate glue droplet [43]. However, the reorganization between the cement-like glue and the pyriform spidroin occurs only near the boundary of these two materials, and it may not generate strong SFG.

For cyclic lactams (candidate 5)) with ring sizes of seven or less in a solid state, the band near 3280 cm$^{-1}$ is replaced by a new one near 3175 cm$^{-1}$ [66,104,113,114]. Darmon *et al.* [104] suggested that the absorption near 3280 cm$^{-1}$ is due to hydrogen bonds between amides in the *trans*-configuration, whereas the absorption near 3160 cm$^{-1}$ arises from ones in the *cis*-configurations. However, the cyclic lactam is unlikely to form an oriented cluster in pyriform silk, and it may not be the origin of the 3150 cm$^{-1}$ peak.

In Fig. 4(b), the relative SFG intensity at 3150 cm$^{-1}$ is almost the same between SSP and PPP polarization combination. Since this mode has a quite different polarization dependence from those of the NH mode at 3270 cm$^{-1}$ and 3300 cm$^{-1}$, the molecular vibration at 3150 cm$^{-1}$ should be oriented statistically in a different direction from the average NH bonds in the whole attachment disc. The orientation of NH and/or NH$_2$ at the amino acid residue may have some effect on the interaction between the pyriform silk and the cement-like glue. However, since interfacial vibration between NH and/or NH$_2$ groups and cement-like glue was not directly observed, we cannot say anything concerning the molecular mechanism of the adhesion property.

## 5. Summary and Conclusions

SFG spectra of spider silk in the attachment disc have been obtained in the hydrogen stretching vibration region between 2800 and 3500 cm$^{-1}$. Doublet peaks of the NH band due to the peptide component of the attachment disc were observed in the SFG spectrum at 3250-3320 cm$^{-1}$. This doublet can be attributed to two different NH vibration modes. The doublet peaks can be assigned to the α-helix and/or β-sheet structure. By comparing $\chi_{ppp}$ and $\chi_{ssp}$, we concluded that the molecular vibration at 3150 cm$^{-1}$ should be oriented statistically in a different direction from the average NH bonds in the whole attachment disc. This vibration mode is assigned to the NH group of arginine side chain and/or NH$_2$ group of arginine and glutamine side chain. Although the SFG spectrum of the attachment disc and aggregate glue droplet contains common properties, the behavior of 3150 cm$^{-1}$ and 3300 cm$^{-1}$ of the attachment disc is clearly different from that of the aggregate glue droplet. An observation by SFG microscopy will give more information in the future, while the analysis should be upgraded to the one using a DFT method.

**CRediT authorship contribution statement**
**Yue Zhao:** Conceptualization, Visualization, Writing–original draft, Methodology. **Lin Liang:** Investigation, Data curation. **Yanrong Li:** Investigation, Data curation. **Khuat Thi Thu Hien:** Supervision. **Goro Mizutani:** Supervision, Validation. **Harvey N. Rutt:** Validation.

**ORCID:**
Yue Zhao: 0000-0002-8550-2020
Goro Mizutani: 0000-0002-4534-9359

**Acknowledgment**
This research was financially supported by a JAIST Research Grant (Grant for Fundamental Research in 2017) of Japan Advanced Institute of Science and Technology Foundation (JAIST Foundation).

**Appendix A. Supplementary data**
**Video S1.** The detail of sample collection. **Figure S1.** The angular dependence of the specular linear optical reflectivity of the attachment disc. **Figure S2.** The FT-IR spectrum of the 1800-2800 cm-1 region of an attachment disc of a spider (*N. clavata*). **Figure S3.** The second harmonic generation microscopy of the attachment disc.

# References

[1] F.G. Barth, S.N. Gorb, M.A. Landolfa, Dragline-associated behaviour of the orb web spider Nephila clavipes (Araneoidea, Tetragnathidae), J. Zool. 244 (1998) 323–330.

[2] F. Saffre, A.-C. Mailleux, J.-L. Deneubourg, Dragline attachment pattern in the neotropical social spider Anelosimus eximius (Araneae: Theridiidae), J. Insect Behav. 12 (1999) 277–282.

[3] M.A. Townley, E.K. Tillinghast, On the use of ampullate gland silks by wolf spiders (Araneae, Lycosidae) for attaching the egg sac to the spinnerets and a proposal for defining nubbins and tartipores, J. Arachnol. 31 (2003) 209–245.

[4] W.G. Eberhard, Possible functional significance of spigot placement on the spinnerets of spiders, J. Arachnol. (2010) 407–414.

[5] J. Kovoor, L. Zylberberg, Fine structural aspects of silk secretion in a spider (Araneus diadematus). I. Elaboration in the pyriform glands, Tissue Cell. 12 (1980) 547–556.

[6] E. Blasingame, T. Tuton-Blasingame, L. Larkin, A.M. Falick, L. Zhao, J. Fong, V. Vaidyanathan, A. Visperas, P. Geurts, X. Hu, others, Pyriform spidroin 1, a novel member of the silk gene family that anchors dragline silk fibers in attachment discs of the black widow spider, Latrodectus Hesperus, J. Biol. Chem. 284 (2009) 29097–29108.

[7] J.O. Wolff, I. Grawe, M. Wirth, A. Karstedt, S.N. Gorb, Spider's super-glue: thread anchors are composite adhesives with synergistic hierarchical organization, Soft Matter. 11 (2015) 2394–2403.

[8] P. Geurts, L. Zhao, Y. Hsia, E. Gnesa, S. Tang, F. Jeffery, C. La Mattina, A. Franz, L. Larkin, C. Vierra, Synthetic spider silk fibers spun from pyriform spidroin 2, a glue silk protein discovered in orb-weaving spider attachment discs, Biomacromolecules. 11 (2010) 3495–3503.

[9] J. Hajer, D. Reháková, Spinning activity of the spider Trogloneta granulum (Araneae, Mysmenidae): web, cocoon, cocoon handling behaviour, draglines and attachment discs, Zoology. 106 (2003) 223–231.

[10] V. Sahni, J. Harris, T.A. Blackledge, A. Dhinojwala, Cobweb-weaving spiders produce different attachment discs for locomotion and prey capture, Nat. Commun. 3 (2012) 1106.

[11] M.-J. Moon, J.-S. An, Microstructure of the silk apparatus of the comb-footed spider, Achaearanea tepidariorum (Araneae: Theridiidae), Entomol. Res. 36 (2006) 56–63.

[12] J.O. Wolff, M.E. Herberstein, Three-dimensional printing spiders: back-and-forth glue application yields silk anchorages with high pull-off resistance under varying loading situations, J. R. Soc. Interface. 14 (2017) 20160783.

[13] I. Grawe, J.O. Wolff, S.N. Gorb, Composition and substrate-dependent strength of the silken attachment discs in spiders, J. R. Soc. Interface. 11 (2014) 20140477.

[14] A.J. Kinloch, Adhesion and adhesives: science and technology, J. Control. Release. 7 (1987) 288–289.

[15] R.C. Chaw, C.A. Saski, C.Y. Hayashi, Complete gene sequence of spider attachment silk protein (PySp1) reveals novel linker regions and extreme repeat homogenization, Insect Biochem. Mol. Biol. 81 (2017) 80–90.

[16] Y. Zhao, K.T.T. Hien, G. Mizutani, H.N. Rutt, Second-order nonlinear optical microscopy of spider silk, Appl. Phys. B. 123 (2017) 188.

[17] Y. Zhao, Y. Li, K.T.T. Hien, G. Mizutani, H.N. Rutt, Observation of spider silk by femtosecond pulse laser second harmonic generation microscopy, Surf. Interface Anal. 51 (2019) 50–56.

[18] J. Wang, M.A. Even, X. Chen, A.H. Schmaier, J.H. Waite, Z. Chen, Detection of amide I signals of interfacial proteins in situ using SFG, J. Am. Chem. Soc. 125 (2003) 9914–9915.
[19] L. Fu, G. Ma, E.C.Y. Yan, In situ misfolding of human islet amyloid polypeptide at interfaces probed by vibrational sum frequency generation, J. Am. Chem. Soc. 132 (2010) 5405–5412.
[20] L. Fu, Z. Wang, B.T. Psciuk, D. Xiao, V.S. Batista, E.C.Y. Yan, Characterization of parallel β-sheets at interfaces by chiral sum frequency generation spectroscopy, J. Phys. Chem. Lett. 6 (2015) 1310–1315.
[21] D. Xiao, L. Fu, J. Liu, V.S. Batista, E.C.Y. Yan, Amphiphilic adsorption of human islet amyloid polypeptide aggregates to lipid/aqueous interfaces, J. Mol. Biol. 421 (2012) 537–547.
[22] Y. Liu, J. Jasensky, Z. Chen, Molecular interactions of proteins and peptides at interfaces studied by sum frequency generation vibrational spectroscopy, Langmuir. 28 (2012) 2113–2121.
[23] E.C.Y. Yan, L. Fu, Z. Wang, W. Liu, Biological macromolecules at interfaces probed by chiral vibrational sum frequency generation spectroscopy, Chem. Rev. 114 (2014) 8471–8498.
[24] B. Ding, J. Jasensky, Y. Li, Z. Chen, Engineering and characterization of peptides and proteins at surfaces and interfaces: a case study in surface-sensitive vibrational spectroscopy, Acc. Chem. Res. 49 (2016) 1149–1157.
[25] M.F.M. Engel, C.C. VandenAkker, M. Schleeger, K.P. Velikov, G.H. Koenderink, M. Bonn, The polyphenol EGCG inhibits amyloid formation less efficiently at phospholipid interfaces than in bulk solution, J. Am. Chem. Soc. 134 (2012) 14781–14788.
[26] S. Ye, H. Li, W. Yang, Y. Luo, Accurate determination of interfacial protein secondary structure by combining interfacial-sensitive amide I and amide III spectral signals, J. Am. Chem. Soc. 136 (2014) 1206–1209.
[27] D.K. Schach, W. Rock, J. Franz, M. Bonn, S.H. Parekh, T. Weidner, Reversible activation of a cell-penetrating peptide in a membrane environment, J. Am. Chem. Soc. 137 (2015) 12199–12202.
[28] T.W. Golbek, J. Franz, J. Elliott Fowler, K.F. Schilke, T. Weidner, J.E. Baio, Identifying the selectivity of antimicrobial peptides to cell membranes by sum frequency generation spectroscopy, Biointerphases. 12 (2017) 02D406.
[29] T. Weidner, N.T. Samuel, K. McCrea, L.J. Gamble, R.S. Ward, D.G. Castner, Assembly and structure of α-helical peptide films on hydrophobic fluorocarbon surfaces, Biointerphases. 5 (2010) 9–16.
[30] X. Chen, A.P. Boughton, J.J.G. Tesmer, Z. Chen, In situ investigation of heterotrimeric G protein βγ subunit binding and orientation on membrane bilayers, J. Am. Chem. Soc. 129 (2007) 12658–12659.
[31] K.M. Reiser, A.B. McCourt, D.R. Yankelevich, A. Knoesen, Structural origins of chiral second-order optical nonlinearity in collagen: amide I band, Biophys. J. 103 (2012) 2177–2186.
[32] G. Niaura, Z. Kuprionis, I. Ignatjev, M. Kažemėkaitė, G. Valincius, Z. Talaikytė, V. Razumas, A. Svendsen, Probing of lipase activity at air/water interface by sum-frequency generation spectroscopy, J. Phys. Chem. B. 112 (2008) 4094–4101.
[33] M. Okuno, T. Ishibashi, Heterodyne-detected achiral and chiral vibrational sum frequency

generation of proteins at air/water interface, J. Phys. Chem. C. 119 (2015) 9947–9954.

[34] S. Hosseinpour, S.J. Roeters, M. Bonn, W. Peukert, S. Woutersen, T. Weidner, Structure and dynamics of interfacial peptides and proteins from vibrational sum-frequency generation spectroscopy, Chem. Rev. 120 (2020) 3420–3465.

[35] E.C.Y. Yan, Z. Wang, L. Fu, Proteins at interfaces probed by chiral vibrational sum frequency generation spectroscopy, J. Phys. Chem. B. 119 (2015) 2769–2785.

[36] J. Wang, S.-H. Lee, Z. Chen, Quantifying the ordering of adsorbed proteins in situ, J. Phys. Chem. B. 112 (2008) 2281–2290.

[37] S. Ye, K.T. Nguyen, S. V Le Clair, Z. Chen, In situ molecular level studies on membrane related peptides and proteins in real time using sum frequency generation vibrational spectroscopy, J. Struct. Biol. 168 (2009) 61–77.

[38] J. Wang, S.M. Buck, Z. Chen, Sum frequency generation vibrational spectroscopy studies on protein adsorption, J. Phys. Chem. B. 106 (2002) 11666–11672.

[39] J. Wang, M.L. Clarke, X. Chen, M.A. Even, W.C. Johnson, Z. Chen, Molecular studies on protein conformations at polymer/liquid interfaces using sum frequency generation vibrational spectroscopy, Surf. Sci. 587 (2005) 1–11.

[40] J. Tan, J. Zhang, Y. Luo, S. Ye, Misfolding of a Human Islet Amyloid Polypeptide at the Lipid Membrane Populates through β-Sheet Conformers without Involving α-Helical Intermediates, J. Am. Chem. Soc. 141 (2019) 1941–1948.

[41] L. Fu, J. Liu, E.C.Y. Yan, Chiral sum frequency generation spectroscopy for characterizing protein secondary structures at interfaces, J. Am. Chem. Soc. 133 (2011) 8094–8097.

[42] L. Fu, Z. Wang, V.S. Batista, E.C.Y. Yan, New insights from sum frequency generation vibrational spectroscopy into the interactions of islet amyloid polypeptides with lipid membranes, J. Diabetes. Res. 2016 (2016) 7293063.

[43] S. Singla, G. Amarpuri, N. Dhopatkar, T.A. Blackledge, A. Dhinojwala, Hygroscopic compounds in spider aggregate glue remove interfacial water to maintain adhesion in humid conditions, Nat. Commun. 9 (2018) 1890.

[44] E.K. Tillinghast, Selective removal of glycoproteins from the adhesive spiral of the spiders orb web, Naturwissenschaften. 68 (1981) 526–527.

[45] O. Choresh, B. Bayarmagnai, R. V Lewis, Spider web glue: two proteins expressed from opposite strands of the same DNA sequence, Biomacromolecules. 10 (2009) 2852–2856.

[46] M.A. Collin, T.H. Clarke, N.A. Ayoub, C.Y. Hayashi, Evidence from multiple species that spider silk glue component ASG2 is a spidroin, Sci. Rep. 6 (2016) 21589.

[47] K. Dreesbach, G. Uhlenbruck, E.K. Tillinghast, Carbohydrates of the trypsin soluble fraction of the orb web of Argiope trifasciata, Insect Biochem. 13 (1983) 627–631.

[48] E.K. Tillinghast, H. Sinohara, Carbohydrates associated with the orb web protein of Argiope aurantia, Biochem. Int. 9 (1984) 315–317.

[49] E.K. Tillinghast, M.A. Townley, T.N. Wight, G. Uhlenbruck, E. Janssen, The adhesive glycoprotein of the orb web of Argiope aurantia (Araneae, Araneidae), Mater. Res. Soc. Symp. Proc. 292 (1992) 9–23.

[50] E.K. Tillinghast, The chemical fractionation of the orb web of Argiope spiders, Insect Biochem. 14 (1984) 115–120.

[51] F. Vollrath, W.J. Fairbrother, R.J.P. Williams, E.K. Tillinghast, D.T. Bernstein, K.S. Gallagher, M.A. Townley, Compounds in the droplets of the orb spider's viscid spiral, Nature. 345 (1990) 526–528.

[52] D. Jain, G. Amarpuri, J. Fitch, T.A. Blackledge, A. Dhinojwala, Role of Hygroscopic Low Molecular Mass Compounds in Humidity Responsive Adhesion of Spider's Capture Silk, Biomacromolecules. 19 (2018) 3048–3057.

[53] H. Schildknecht, P. Kunzelmann, D. Krauss, C. Kuhn, Uber die Chemie der Spinnwebe, I Arthropodenabwehrstoffe, LVII, Naturwissenschaften. 59 (1972) 98–99.

[54] E.K. Tillinghast, T. Christenson, Observations on the chemical composition of the web of Nephila clavipes (Araneae, Araneidae), J. Arachnol. 12 (1984) 69–74.

[55] M.A. Townley, D.T. Bernstein, K.S. Gallagher, E.K. Tillinghast, Comparative study of orb web hygroscopicity and adhesive spiral composition in three araneid spiders, J. Exp. Zool. 259 (1991) 154–165.

[56] Y. Zhao, M. Morita, T. Sakamoto, Analysis the water in aggregate glue droplets of spider orb web by TOF-SIMS, Surf. Interface Anal. 53 (2021) 359–364.

[57] P. Papadopoulos, J. Sölter, F. Kremer, Structure-property relationships in major ampullate spider silk as deduced from polarized FTIR spectroscopy, Eur. Phys. J. E. 24 (2007) 193–199.

[58] P.R. Laity, S.E. Gilks, C. Holland, Rheological behaviour of native silk feedstocks, Polymer (Guildf). 67 (2015) 28–39.

[59] A.A. Dzhatdoeva, A.M. Polimova, E. V Proskurnina, M.A. Proskurnin, Y.A. Vladimirov, Determination of lipids and their oxidation products by IR spectrometry, J. Anal. Chem. 71 (2016) 542–548.

[60] A. Blume, Properties of lipid vesicles: FT-IR spectroscopy and fluorescence probe studies, Curr. Opin. Colloid Interface Sci. 1 (1996) 64–77.

[61] A. Savitzky, M.J.E. Golay, Smoothing and differentiation of data by simplified least squares procedures., Anal. Chem. 36 (1964) 1627–1639.

[62] M. Rozenberg, G. Shoham, FTIR spectra of solid poly-l-lysine in the stretching NH mode range, Biophys. Chem. 125 (2007) 166–171.

[63] A. Tinti, M. Di Foggia, P. Taddei, A. Torreggiani, M. Dettin, C. Fagnano, Vibrational study of auto-assembling oligopeptides for biomedical applications, J. Raman Spectrosc. 39 (2008) 250–259.

[64] D. Ke, C. Zhan, X. Li, X. Wang, Y. Zeng, J. Yao, Ultrasound-induced modulations of tetrapeptide hierarchical 1-D self-assembly and underlying molecular structures via sonocrystallization, J. Colloid Interface Sci. 337 (2009) 54–60.

[65] S.-H. Lee, N.G. Mirkin, S. Krimm, A quantitative anharmonic analysis of the amide A band in α-helical poly (L-alanine), Biopolymers. 49 (1999) 195–207.

[66] L.J. Bellamy, Amides, proteins and polypeptides, in: Infra-Red Spectra Complex Mol., Springer, 1975: pp. 231–262.

[67] M. Tsuboi, T. Shimanouchi, S.-I. Mizushima, Near Infrared Spectra of Compounds with Two Peptide Bonds and the Configuration of a Polypeptide Chain. VII. On the Extended Forms of Polypeptide Chains, J. Am. Chem. Soc. 81 (1959) 1406–1411.

[68] T. Weidner, N.F. Breen, G.P. Drobny, D.G. Castner, Amide or amine: Determining the origin of the 3300 cm-1 NH mode in protein SFG spectra using 15N isotope labels, J. Phys. Chem. B. 113 (2009) 15423–15426.

[69] L. Fu, Z. Wang, E.C.Y. Yan, N-H Stretching Modes Around 3300 Wavenumber From Peptide Backbones Observed by Chiral Sum Frequency Generation Vibrational Spectroscopy, Chirality. 26 (2014) 521–524.

[70] M.A. Belkin, T.A. Kulakov, K.-H. Ernst, L. Yan, Y.R. Shen, Sum-frequency vibrational

spectroscopy on chiral liquids: a novel technique to probe molecular chirality, Phys. Rev. Lett. 85 (2000) 4474–4477.

[71] Y.R. Shen, Surface studies by optical second harmonic generation: An overview, J. Vac. Sci. Technol. B. 3 (1985) 1464–1466.

[72] G.L. Richmond, Molecular bonding and interactions at aqueous surfaces as probed by vibrational sum frequency spectroscopy, Chem. Rev. 102 (2002) 2693–2724.

[73] P. Guyot-Sionnest, J.H. Hunt, Y.R. Shen, Sum-frequency vibrational spectroscopy of a Langmuir film: Study of molecular orientation of a two-dimensional system, Phys. Rev. Lett. 59 (1987) 1597–1600.

[74] R.A. MacPhail, H.L. Strauss, R.G. Snyder, C.A. Elliger, Carbon-hydrogen stretching modes and the structure of n-alkyl chains. 2. Long, all-trans chains, J. Phys. Chem. 88 (1984) 334–341.

[75] R. Mendelsohn, J.W. Brauner, A. Gericke, External infrared reflection absorption spectrometry of monolayer films at the air-water interface, Annu. Rev. Phys. Chem. 46 (1995) 305–334.

[76] Z. Chen, R. Ward, Y. Tian, F. Malizia, D.H. Gracias, Y.R. Shen, G.A. Somorjai, Interaction of fibrinogen with surfaces of end-group-modified polyurethanes: A surface-specific sum-frequency-generation vibrational spectroscopy study, J. Biomed. Mater. Res. 62 (2002) 254–264.

[77] J.C. Goodwin, L. Tjarks, Preparation and Structure of (E)-1-(3′-Hydroxy-2-Furanyl)-3-(3″-Hydroxy-4″-Methoxyphenyl)-2-Propen-1-One, J. Carbohydr. Chem. 7 (1988) 133–140.

[78] M. Xu, S. Celerier, J.-D. Comparot, J. Rousseau, M. Corbet, F. Richard, J.-M. Clacens, Upgrading of furfural to biofuel precursors via aldol condensation with acetone over magnesium hydroxide fluorides $MgF_{2-x}(OH)_x$, Catal. Sci. Technol. 9 (2019) 5793–5802.

[79] S.S. Pandey, W. Takashima, K. Kaneto, Role of processing solvents in photocarrier transport of polyanilines, Jpn. J. Appl. Phys. 39 (2000) 4045.

[80] K.S. Fathima, M. Sathiyendran, K. Anitha, Structure Elucidation, Biological Evaluation and Molecular Docking Studies on a new organic salt: 2-aminobenzimidazolium 5-nitro-2-hydroxybenzoate, J. Mol. Struct. 1177 (2019) 457–468.

[81] G. Varsányi, Assignments for vibrational spectra of seven hundred benzene derivatives, Hilger, London, 1974.

[82] Y. Yu, N. Chu, Q. Pan, M. Zhou, S. Qiao, Y. Zhao, C. Wang, X. Li, Solvent Effects on Gelation Behavior of the Organogelator Based on L-Phenylalanine Dihydrazide Derivatives, Materials (Basel). 12 (2019) 1890.

[83] D. Scuderi, J.M. Bakker, S. Durand, P. Maitre, A. Sharma, J.K. Martens, E. Nicol, C. Clavaguera, G. Ohanessian, Structure of singly hydrated, protonated phospho-tyrosine, Int. J. Mass Spectrom. 308 (2011) 338–347.

[84] M. Marinescu, D.G. Tudorache, G.I. Marton, C.-M. Zalaru, M. Popa, M.-C. Chifiriuc, C.-E. Stavarache, C. Constantinescu, Density functional theory molecular modeling, chemical synthesis, and antimicrobial behaviour of selected benzimidazole derivatives, J. Mol. Struct. 1130 (2017) 463–471.

[85] J.L. Castro, M.R.L. Ramirez, J.F. Arenas, J.C. Otero, Vibrational spectra of 3-phenylpropionic acid and L-phenylalanine, J. Mol. Struct. 744 (2005) 887–891.

[86] R. Linder, M. Nispel, T. Häber, K. Kleinermanns, Gas-phase FT-IR-spectra of natural

amino acids, Chem. Phys. Lett. 409 (2005) 260–264.

[87] N. Hammond, Retraction: The next generation cell-penetrating peptide and carbon dot conjugated nano-liposome for transdermal delivery of curcumin, Biomater. Sci. 7 (2018) 442.

[88] T. Kolev, B.B. Koleva, E. Cherneva, M. Spiteller, M. V Winter, W.S. Sheldrick, H. Mayer-Figge, Crystal structure, IR-LD spectroscopic, theoretical and vibrational analysis of valinamide ester amide of squaric acid diethyl ester, Struct. Chem. 17 (2006) 491–499.

[89] C.M. Lee, K. Kafle, Y.B. Park, S.H. Kim, Probing crystal structure and mesoscale assembly of cellulose microfibrils in plant cell walls, tunicate tests, and bacterial films using vibrational sum frequency generation (SFG) spectroscopy, Phys. Chem. Chem. Phys. 16 (2014) 10844–10853.

[90] Y.B. Park, K. Kafle, C.M. Lee, D.J. Cosgrove, S.H. Kim, Does cellulose II exist in native alga cell walls? Cellulose structure of Derbesia cell walls studied with SFG, IR and XRD, Cellulose. 22 (2015) 3531–3540.

[91] S. Olejnik, A.M. Posner, J.P. Quirk, The IR spectra of interlamellar kaolinite-amide complexes, I. The complexes of formamide, N-methylformamide and dimethylformamide, Clays Clay Miner. 19 (1971) 83–94.

[92] J.A. Lercher, H. Noller, Infrared spectroscopic study of hydroxyl group acid strength of silica, alumina, and magnesia mixed oxides, J. Catal. 77 (1982) 152–158.

[93] K. Kinoshita, S. Takenaka, M. Hayashi, Isolation of proposed intermediates in the biosynthesis of mycinamicins, Chem. Commun. (1988) 943–945.

[94] C.-H. Lin, H.-S. Chang, H.-R. Liao, I.-S. Chen, I.-L. Tsai, Triterpenoids from the Roots of Rhaphiolepis indica var. tashiroi and Their Anti-Inflammatory Activity, Int. J. Mol. Sci. 14 (2013) 8890–8898.

[95] A.W. Burgess, H.A. Scheraga, Stable conformations of dipeptides, Biopolymers. 12 (1973) 2177–2183.

[96] R.R. Gardner, G.-B. Liang, S.H. Gellman, An Achiral Dipeptide Mimetic That Promotes. beta.-Hairpin Formation, J. Am. Chem. Soc. 117 (1995) 3280–3281.

[97] C.M. Lee, A. Mittal, A.L. Barnette, K. Kafle, Y.B. Park, H. Shin, D.K. Johnson, S. Park, S.H. Kim, Cellulose polymorphism study with sum-frequency-generation (SFG) vibration spectroscopy: identification of exocyclic CH2OH conformation and chain orientation, Cellulose. 20 (2013) 991–1000.

[98] D.J. Perry, D. Bittencourt, J. Siltberg-Liberles, E.L. Rech, R. V Lewis, Piriform spider silk sequences reveal unique repetitive elements, Biomacromolecules. 11 (2010) 3000–3006.

[99] T. Miyazawa, Perturbation treatment of the characteristic vibrations of polypeptide chains in various configurations, J. Chem. Phys. 32 (1960) 1647–1652.

[100] Z. Movasaghi, S. Rehman, I. Rehman, Fourier transform infrared (FTIR) spectroscopy of biological tissues, Appl. Spectrosc. Rev. 43 (2008) 134–179.

[101] H. Morita, I. Machida, Y. Hirasawa, J. Kobayashi, Taxezopidines M and N, Taxoids from the Japanese Yew, Taxus cuspidata, J. Nat. Prod. 68 (2005) 935–937.

[102] P. Stanley, Golgi glycosylation, Cold Spring Harb. Perspect. Biol. 3 (2011) a005199.

[103] R.E. Richards, H.W. Thompson, 237. Spectroscopic studies of the amide linkage, J. Chem. Soc. (1947) 1248–1260.

[104] S.E. Darmon, G. Sutherland, Evidence from Infra-Red Spectroscopy on the Structure of Proteins, Nature. 164 (1949) 440–441.

[105] H. Letaw Jr, A.H. Gropp, A Study of the Infrared Spectrum of the Amide Group, J. Chem.

Phys. 21 (1953) 1621–1627.

[106] S.-H. Lee, S. Krimm, Polarized Raman spectra of oriented films of α-helical poly (L-alanine) and its N-deuterated analogue, J. Raman Spectrosc. 29 (1998) 73–80.

[107] S. Krimm, J. Bandekar, Vibrational spectroscopy and conformation of peptides, polypeptides, and proteins, Adv. Protein Chem. 38 (1986) 181–364.

[108] S. Kumar, S.B. Rai, Spectroscopic studies of L-arginine molecule, Indian J. Pure Appl. Phys. 48 (2010) 251–255.

[109] M. Rozenberg, G. Shoham, I. Reva, R. Fausto, A correlation between the proton stretching vibration red shift and the hydrogen bond length in polycrystalline amino acids and peptides, Phys. Chem. Chem. Phys. 7 (2005) 2376–2383.

[110] B. Boeckx, G. Maes, The conformational behavior and H-bond structure of asparagine: A theoretical and experimental matrix-isolation FT-IR study, Biophys. Chem. 165 (2012) 62–73.

[111] A. Ebrahiminezhad, Y. Ghasemi, S. Rasoul-Amini, J. Barar, S. Davaran, Impact of amino-acid coating on the synthesis and characteristics of iron-oxide nanoparticles (IONs), Bull. Korean Chem. Soc. 33 (2012) 3957–3962.

[112] L. Dreesen, Y. Sartenaer, C. Humbert, A.A. Mani, C. Méthivier, C.-M. Pradier, P.A. Thiry, A. Peremans, Probing Ligand--Protein Recognition with Sum-Frequency Generation Spectroscopy: The Avidin--Biocytin Case, ChemPhysChem. 5 (2004) 1719–1725.

[113] R. Huisgen, H. Brade, H. Walz, I. Glogger, Mittlere Ringe, VII. Die Eigenschaften Aliphatischer Lactame und Die cis-trans-Isomerie der Säureamidgruppe, Chem. Ber. 90 (1957) 1437–1447.

[114] H.E. Hallam, C.M. Jones, Structures of cyclic amides: Part 2. Associated species, J. Mol. Struct. 1 (1968) 425–435.

# Sum frequency generation spectroscopy of the attachment disc of a spider

*Yue Zhao [1], Lin Liang [1], Yanrong Li [1], Hien Thi Thu Khuat [1], Goro Mizutani [1,*], and Harvey N. Rutt [2]*

[1] School of Materials Science, Japan Advanced Institute of Science and Technology 1-1 Nomi, 923-1292, Japan

[2] School of Electronic and Computer Science, University of Southampton, SO17 1BJ, UK

* E-mail: mizutani@jaist.ac.jp

1. **Video S1** shows the detail of sample collection.

2. **Figure S1** shows the angular dependence and polarization dependence of the specular linear optical reflectivity of the attachment disc.

3. **Figure S2** shows the FT-IR spectrum of the 1800-2800 cm$^{-1}$ region of an attachment disc of a spider (*N. clavata*).

4. **Figure S3** shows the second harmonic generation microscopy of the attachment disc.

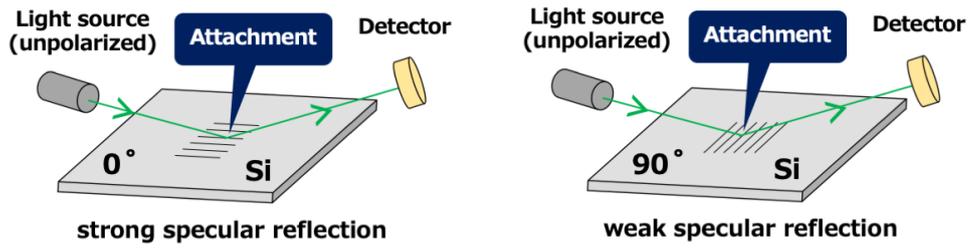

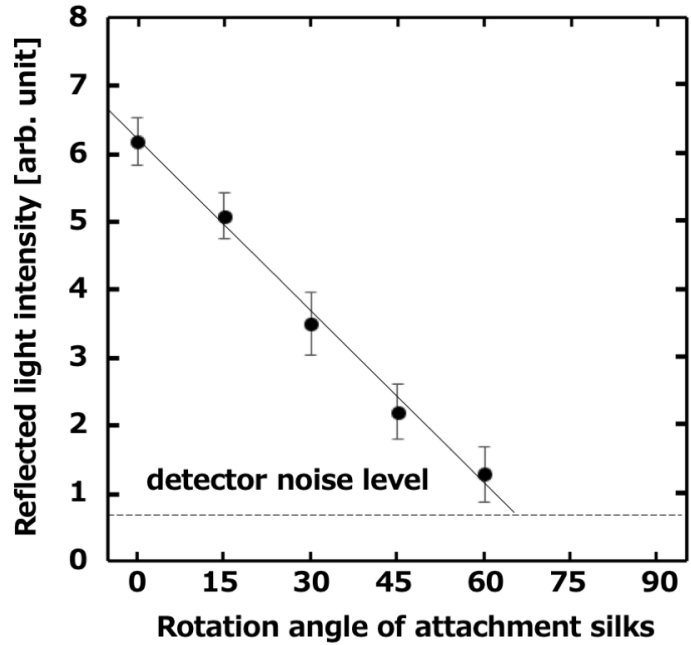

**Figure S1.** Linear light reflection intensity as a function of the rotation angle of the attachment silk at the wavelength of 532 ± 5 nm. 0° is defined as the rotational angle when the attachment silk fiber axis is parallel to the incident plane of light. The intensity of the reflected light was below the noise level above the rotational angle of 70°.

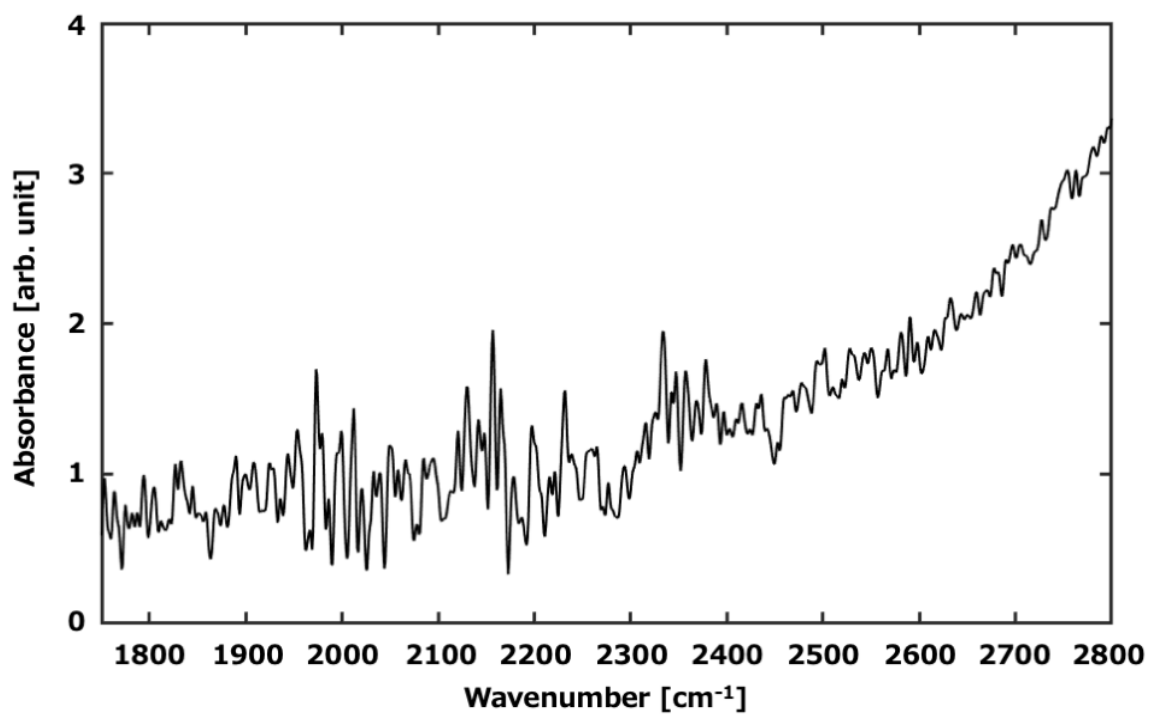

**Figure S2.** FT-IR spectrum of the 1800-2800 cm$^{-1}$ region of an attachment disc of a spider (*N. clavata*).

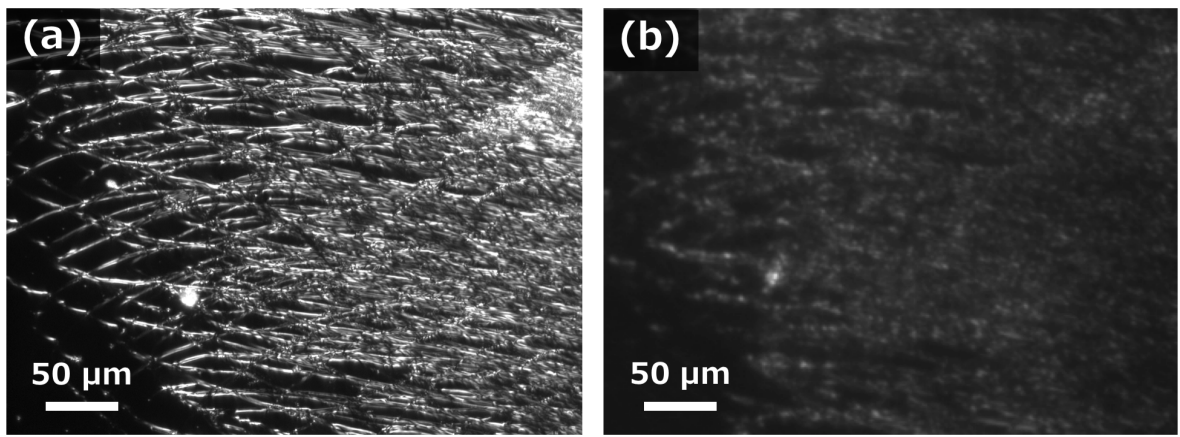

**Figure S3.** (a) Microscopic image of an attachment disc silk fibers taken by a CMOS camera with white light illumination. (b) Second harmonic generation (SHG) image. For SHG image, the excitation light pulses were with a repetition frequency of 1 kHz, wavelength of 800 nm, and a pulse width of about 120 fs. The energy density on sample was 255 μJ/cm$^2$ per pulse. The integration time was 1 s.